\documentclass[a4paper,twocolumn,11pt,accepted=2020-06-19]{quantumarticle}
\pdfoutput=1
\usepackage[utf8]{inputenc}
\usepackage[english]{babel}
\usepackage[T1]{fontenc}
\usepackage{amsmath}
\usepackage{amsfonts}
\usepackage{amssymb}
\usepackage{hyperref}
\usepackage{graphicx}
\usepackage{epsfig,amssymb,amsmath}
\usepackage[noadjust]{cite}

\def\>{\rangle}
\def\<{\langle}

\def\labell#1{\label{#1}}
\def\togli#1{}

\begin{document}

\title{Squeezing Metrology: a unified framework}

\author{Lorenzo Maccone}
\email{maccone@unipv.it}
\author{Alberto Riccardi}
\affiliation{{Dip.~Fisica and INFN Sez.\ Pavia, University of Pavia, via
    Bassi 6, I-27100 Pavia, Italy}}
\email{alberto.riccardi01@universitadipavia.it}
\maketitle

\begin{abstract}
  Quantum metrology theory has up to now focused on the resolution
  gains obtainable thanks to the entanglement among $N$ probes.
  Typically, a quadratic gain in resolution is achievable, going from
  the $1/\sqrt{N}$ of the central limit theorem to the $1/N$ of the
  Heisenberg bound. Here we focus instead on quantum squeezing and
  provide a unified framework for metrology with squeezing, showing
  that, similarly, one can generally attain a quadratic gain when
  comparing the resolution achievable by a squeezed probe to the best
  $N$-probe classical strategy achievable with the same energy.
  Namely, here we give a quantification of the Heisenberg squeezing
  bound for arbitrary estimation strategies that employ squeezing. Our
  theory recovers known results (e.g.~in quantum optics and spin
  squeezing), but it uses the general theory of squeezing and holds
  for arbitrary quantum systems.
\end{abstract}

\section{Introduction}
\begin{figure}[hbt]
\begin{center}
\includegraphics[width=0.45\textwidth]{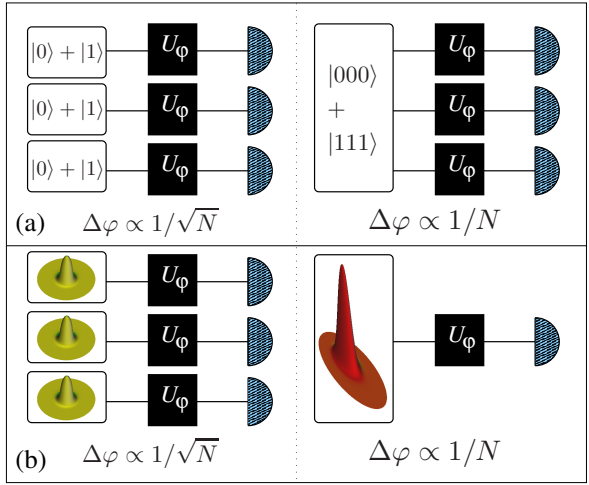}
\end{center}
\vspace{-.5cm}
\caption{Conventional quantum metrology vs.~squeezed metrology.
  (a)~Typically one considers entanglement among probes as a means of
  achieving a quadratic enhancement in precision: $N$ entangled probes
  allow a precision gain from $\Delta\varphi/\sqrt{N}$ (left) to
  $\Delta\varphi/N$ (right) \cite{qmetr}. (b)~Here we consider the
  role of squeezing: if a probe in a classical (coherent) state allows
  a precision estimate of $\Delta\varphi$, one can achieve
  $\Delta\varphi/\sqrt{N}$ by using $N$ such probes (left). Devoting
  the same energy of the $N$ coherent probes that optimally use the
  energy resources to squeezing a single probe, one can decrease the
  error to $\Delta\varphi/N$ (right), a quadratic gain.
  \label{f:comp}}\end{figure}

In quantum metrology one studies the resolution gains that can be
attained when using quantum effects in the estimation strategy. The
usual setting considers an estimation strategy where a parameter
$\varphi$ is encoded onto a probe state through a unitary
transformation $U_\varphi=e^{iH\varphi}$, where $H$ is the probe
Hamiltonian. This is a very general setting that encompasses most
estimations. While squeezing has been used in quantum metrology for
specific systems
\cite{yuen,yuen1,perelomov,ozawa,altri,spinsq,spinsq1,spinsq2,spinsq3,esperim,esperim1,esperim2,lund,ref},
there is no general theory of squeezing-based metrology that holds for
arbitrary measurements and systems. In the general case, quantum
metrology theory
\cite{qmetr,caves,caves1,kok,review,review1,matteo,rafalreview,geza,walm,book,gabriel}
focuses on entanglement: it analyzes the situation in which the
estimation is repeated $N$ times and shows that there is a quadratic
improvement in resolution whenever the probes are entangled. Without
entanglement, one can only achieve the standard-quantum-limit
resolution $\Delta\varphi\propto1/\sqrt{N}$ of the central limit
theorem, and the error decreases to $\Delta\varphi\propto1/N$ of the
Heisenberg bound using entangled probes
\cite{qmetr,review,review1,rafalreview}, Fig.~\ref{f:comp}a.

In this paper we focus on squeezing and consider the case in which a
single probe is squeezed with respect to the observable $A$ that is
measured to estimate the parameter $\varphi$.  We are {\bf not}
claiming to have discovered that squeezing is useful for quantum
metrology (there is plenty of evidence for that in the literature
\cite{yuen,yuen1,perelomov,ozawa,altri,spinsq,spinsq1,spinsq2,spinsq3,esperim,esperim1,esperim2,lund,ref}).
The main result of this paper is a unified framework that describes
all previous (and presumably future) metrology protocols that use
squeezing in {\em any} quantum system and {\em any} observable, as it
is based on the elegant general theory of squeezing for arbitrary
systems \cite{trifonov}. As shown through various examples, the
previously known results
\cite{yuen,yuen1,perelomov,ozawa,altri,spinsq,spinsq1,spinsq2,spinsq3,esperim,esperim1,esperim2,lund,ref}
can be recovered as specific instances of our theory.  Squeezing the
probe requires an amount of energy $E=\<s|H|s\>$, where $|s\>$ is the
squeezed state of the probe.  As in entanglement-based quantum
metrology, a quadratic resolution gain is obtained also in this case,
if one compares the resolution attainable with a squeezed probe to the
resolution obtainable with $N$ classical probes (i.e.~prepared in a
coherent state) of total energy $E$, Fig.~\ref{f:comp}b. By classical
probe we intend a probe prepared in a minimum uncertainty coherent
state, as is customary. We employ $N$ classical probes with the same
energy as the squeezed probe because we need a dimensionless parameter
($N$) to measure the precision enhancement. We compare the quantum and
classical strategies with same average energy because, without any
energy restriction, one could achieve arbitrary precision by using
arbitrarily high squeezing.  In essence, our result can be summarized
as follows: whereas one needs $N$ classical probes to decrease the
error from $\Delta\varphi$ to $\Delta\varphi/\sqrt{N}$, a single
squeezed probe that uses the same energy can decrease the error to
$\Delta\varphi/N$. We show that this resolution increase is the
optimal one (no further enhancement is possible), so we can call it
the Heisenberg squeezing bound. We will be neglecting multiplicative
constants of order one and consider only the scaling in $N$: it is
impossible to give a general theory of the multiplicative constants
because these depend on the specific implementations (as is clear from
the examples below).

The paper outline: we prove the general theory of squeezing metrology,
giving the yields of the possible strategies that employ squeezing;
we then illustrate it with some prototypical cases:
quadrature squeezing, interferometric phase estimation, and spin
squeezing. We conclude with a step-by-step procedure to obtain new
squeezing based metrology protocols.

\section{Results}
Two observables are relevant in an estimation problem: the observable
$A$ that is measured and whose outcomes are used to estimate the
parameter $\varphi$ and the Hamiltonian $H$ that encodes the parameter
onto the probes, namely the generator of translations of $\varphi$
through $U_\varphi=e^{iH\varphi}$. It is then natural to consider as
``classical'' the strategy that uses the coherent states for these two
observables \cite{trifonov,perelomov}. Then the ``quantum'' strategies
are the ones where we squeeze $A$ and anti-squeeze $H$. Squeezed
states for two observables $A$ and $H$ are the eigenstates of the
operator $L(\lambda)\equiv\lambda A+iH$ \cite{trifonov}, where
$\lambda\in{\mathbb{C}}$ is a squeezing parameter: the states are
squeezed in $A$ and antisqueezed in $H$ for $|\lambda|>1$ and squeezed
in $H$ and antisqueezed in $A$ for $|\lambda|<1$ \footnote{In the case
  in which $i[A,B]$ is not strictly positive or negative, as for spin
  systems, additional squeezed states may exist \cite{trifonov}.}. All
states with $|\lambda|=1$ are coherent states. [The converse is not
true in Perelomov's definition of coherent states \cite{perelomov},
where $|\lambda|\neq 1$ for some. Here we use Trifonov's notation
\cite{trifonov}: all have $|\lambda|=1$.] In the following, among all
eigenstates with $|\lambda|=1$, we will consider the minimum
uncertainty ones: these are the ones that are customarily considered
the ``classical'' states.  These
definitions of squeezed and coherent arise from the Schr\"odinger
uncertainty relation
\begin{align}
\Delta A^2\Delta
H^2\geqslant\;&|\frac12\<[A,H]\>|^2\\&+|\frac12\<\{A,H\}\> 
-\<A\>\<H\>|^2
\labell{schr},
\end{align}
where $\Delta X^2$ is the variance of $X$, $\<X\>$ its expectation
value, and $[\,,]$ and $\{\,,\}$ are the commutator and
anticommutator. For the eigenstates of $L(\lambda)$, \cite{trifonov}
\begin{eqnarray}
  \Delta A^2=\frac{|\<C\>|}{2Re\lambda},\;\Delta
  H^2=|\lambda|^2\frac{|\<C\>|}{2Re\lambda},\;\nonumber\\ \Delta
  AH=-\frac{|\<C\>|Im\lambda}{2Re\lambda},
\labell{trif}\;
\end{eqnarray}
where $C\equiv[A,H]$ and $\Delta
AH\equiv\frac12\<\{A,H\}\>-\<A\>\<H\>$ (and additional dimensionful
constant may be present if $A$ and $H$ have different units). By
considering a real $\lambda$, we can restrict to the
Heisenberg-Robertson inequality \cite{robertson} $\Delta A\Delta
H\geqslant\frac12|\<C\>|$ and squeezed states: the best metrological
advantage is obtained in this case (this can be easily shown by
repeating the derivation below with complex $\lambda$), so we will
consider only real positive $\lambda$.\togli{Vedere gli appunti di
  Alberto per il caso $\lambda\in{\mathbb{C}}$.  Si puo' aggiungere in
  appendice?}

When using the outcomes of $A$ to estimate the parameter $\varphi$,
error propagation implies $\Delta\varphi=\Delta
A/|\frac\partial{\partial\varphi}\<A\>|$. We can evaluate the
derivative recalling that the state of the probe is evolved by the
unitary $U_\varphi=e^{iH\varphi}$, so that 
\begin{eqnarray}
\tfrac\partial{\partial\varphi}\<A\>=\tfrac\partial{\partial\varphi}
\<0|e^{-iH\varphi}Ae^{iH\varphi}|0\>=-i\<[H,A]\>
\labell{aggi}\;,
\end{eqnarray}
where $|0\>$ is the initial state of the probe.  So
\begin{eqnarray}
\Delta\varphi=\Delta A/|\<C\>|=1/\sqrt{2\lambda|\<C\>|}
\labell{sq1}\;,
\end{eqnarray}
where we used (\ref{trif}). This equation is also valid for coherent
states where $\lambda=1$. When repeating the estimation procedure $N$
times, the central limit theorem implies that the rmse
reduces by $\sqrt{N}$ to
$\Delta\varphi(N)\equiv\Delta\varphi/\sqrt{N}$.

We now show that a squeezed state with the same energy as $N$
minimum-uncertainty coherent states can allow a quadratic gain when
comparing $\Delta\varphi_{sq}$ to
$\Delta\varphi_{cl}(N)=\Delta\varphi_{cl}/\sqrt{N}$ (i.e.~what can be
achieved classically by repeating $N$ times), where the suffix $sq$
and $cl$ indicates that the respective quantities are calculated on
squeezed or minimum-uncertainty coherent states. Namely,
$\Delta\varphi_{sq}/\Delta\varphi_{cl}\sim 1/N$ which is a quadratic
gain over $\Delta\varphi_{cl}(N)/\Delta\varphi_{cl}=1/\sqrt{N}$. Here
$N$ is the number of coherent probes that can be produced with the
squeezed state's energy, namely
\begin{eqnarray}
N=(\<H\>_{sq}-E_0)/(\<H\>_{cl}-E_0)
\labell{enne}\;,
\end{eqnarray}where $E_0$ is the ground state energy. Clearly we are
interested in comparing the best estimation strategies, since one can
always get a worse strategy by simply wasting energy resources. 

For arbitrary settings, not just the setup here, the best estimation
strategies are the ones that satisfy the quantum measurement bound of
\cite{heisbound} with equality. This bound implies that
\begin{eqnarray}
\Delta\varphi\geqslant\max\Big[\frac\kappa{\nu(\<H\>-E_0)},
\frac\gamma{\sqrt{\nu}\Delta  H}\Big]
\labell{heisb}\;,
\end{eqnarray}
where $\kappa$ and $\gamma$ are numerical constants of order 1 (see
\cite{heisbound}) and $\nu$ is the number of times the estimation is
repeated (here we will consider $\nu=1$). The first term in the max
expresses a quantum speed limit \cite{margolus} and the second arises
from the time-energy uncertainty \cite{man} (our choice $\nu=1$
follows from these results \cite{heisbound}). Eq.~(\ref{heisb})
implies that any energy beyond the standard deviation will be wasted
for the estimation: if $\<H\>-E_0>\zeta\Delta H$ (with
$\zeta\equiv\kappa/\gamma$), then the error $\Delta\varphi$ is
dominated by $\Delta H$: ``too much energy'' strategies. To avoid
wastes, we should choose $\<H\>-E_0\simeq\zeta\Delta H$ (``good
strategies''), where the ``$\simeq$'' sign emphasizes that only the
order of magnitude of the two terms is important since
Eq.~(\ref{heisb}) is not a tight bound \cite{heisbound}.
Eq.~(\ref{heisb}) also implies that estimation strategies that have
$\<H\>-E_0<\zeta\Delta H$ (``too little energy'' strategy) have error
$\Delta\varphi$ dominated by the energy: they cannot achieve the error
of Eq.~(\ref{sq1}), but are limited to
$\Delta\varphi\sim\kappa/(\<H\>-E_o)$.

The ``good strategy'' energy requirement $\<H\>-E_0\simeq\zeta\Delta
H$ can always be enforced when squeezing is used, since the ratio
between energy and its standard deviation can be tuned through the
squeezing parameter $\lambda$: one can assign an arbitrary fraction of
the energy budget to squeezing the probe. Instead, for classical
strategies, the requirement of having a coherent state may be
inconsistent with good strategies: for example spin coherent states
have ``too much energy''.  Otherwise typically the ``good strategy''
requirement can be satisfied only for a specific value of the energy,
as for the harmonic oscillator. Indeed, Eq.~(\ref{trif}) fixes $\Delta
H^2$ to ${|\<C\>|/2}$, and fixing it also to $\<H-E_0\>^2$ may be
possible for coherent states only for a given energy. For classical
strategies with too little energy, using (\ref{enne}) will
underestimate $N$ since some of the coherent state energy is not used
for estimation, see Eq.~(\ref{heisb}). For these strategies, one
should use $N=(\<H\>_{sq}-E_0)/\Delta H_{cl}$ in place of (\ref{enne})
to count only for the energy that is actually employed for the
estimation in classical strategies. The focus on ``good strategies''
is not a limitation of our paper, because these are the strategies
that are optimal: the ones that do not waste energy.

Consider the ``good strategies'' first. In this case, replacing
$\<H\>-E_0=\zeta\Delta H$ into (\ref{enne}) we find
\begin{eqnarray}
N\simeq\frac{\Delta H_{sq}}{\Delta
H_{cl}}=\sqrt{\lambda\left|{\<C\>_{sq}}/{\<C\>_{cl}}\right|}
\labell{enne2}\;,
\end{eqnarray}
where we used (\ref{trif}) in the last equality. Using (\ref{sq1}), we
find
\begin{eqnarray}
\frac{\Delta\varphi_{sq}}{\Delta\varphi_{cl}}=
\sqrt{\left|{\<C\>_{cl}}/{\lambda\<C\>_{sq}}\right|}=\frac 1{N}
\labell{sq2}\;,
\end{eqnarray}
a quadratic gain over the classical strategy of repeating $N$ times
the classical estimation:
$\Delta\varphi_{cl}(N)/\Delta\varphi_{cl}=1/\sqrt{N}$. In other words,
when comparing the squeezed and classical strategy {\em of equal
  energy}, we find
$\Delta\varphi_{sq}/\Delta\varphi_{cl}(N)=1/\sqrt{N}$, the main result
of our paper. Since we obtained the quadratic gain when comparing the
best estimation strategies, this is optimal: one cannot achieve a
larger gain (unless one employs suboptimal classical strategies). This
justifies our claim that (\ref{sq2}) represents the Heisenberg
squeezing bound.

We emphasize that in this derivation we have neglected multiplicative
and additive constant factors of order one: as a general theory
containing them is impossible (they depend on the specific
implementations). In principle, in the abstract scenario we analyzed
here, one may consider a {\em single} coherent state that is not
minimum-uncertainty (namely a state with equal spreads
$\Delta A=\Delta H$ which are larger than their minimum value) and
devote all the energy to it. In this case, that state would already
achieve the optimal precision without any need for squeezing. However,
this strategy does not exist in all the examples we analyzed because
those coherent states have always a bounded variance and the ``good
strategy'' condition cannot be met. Moreover, these large-uncertainty
coherent states would not be recognized as a ``classical'' strategy.

Consider now bad estimation strategies. For classical strategies with
too much energy, the above results still hold if one uses the
appropriate $N$ discussed above. For strategies with too little
energy, Eq.~(\ref{sq1}) does not apply and we need to amend the above
derivation, but we still obtain the same result: consider first the
case of a suboptimal classical estimation strategy, where
$\Delta\varphi_{cl}\sim\kappa/(\<H\>_{cl}-E_o)$, so
\begin{eqnarray} \frac{\Delta\varphi_{sq}}{\Delta\varphi_{cl}}=
\frac{\<H\>_{cl}-E_0}{\kappa\sqrt{2\lambda|\<C\>_{sq}|}}
=\frac1{2\kappa}\frac1{N}
\labell{sq3}\;,
\end{eqnarray}
where we used the fact that $N=\Delta H_{sq}/(\<H\>_{cl}-E_0)$ in this
case. Again we find a quadratic gain (apart from an inconsequential
numerical constant of order one). We obtain analogous results when
comparing two bad estimation strategies or a bad squeezed strategy
with a good classical strategy.

\section{Discussion} (i)~The above proof elucidates why squeezing is
beneficial: only a squeezed estimation strategy can attain the
equality $\<H\>-E_0=\zeta\Delta H$ for all energies. A coherent state
typically attains this only for a specific energy, so it is suboptimal
when the procedure is repeated $N$ times: the $\<H\>$ term in
(\ref{heisb}) scales linearly with $\nu=N$, whereas the $\Delta H$
term scales as $\sqrt{\nu}=\sqrt{N}$.  Hence, given an energy budget,
repeating the measurement is never advantageous and a single-shot {\em
  classical} strategy that optimally uses all resources may exist only
for a specific energy.  Then, only for this value of the energy, the
single-shot classical strategy performs as well as the squeezed one,
otherwise squeezing the probe is always better. (ii)~The squeezing
parameter $\lambda$ is not present in the final result of (\ref{sq2}).
Nonetheless, the results obtained hold only for $\lambda>1$,
corresponding to a reduction in the fluctuations of the observable $A$
that is measured, as expected. Indeed, for $\lambda<1$, corresponding
to an increase in fluctuations of $A$, Eq.~(\ref{enne2}) typically
implies that $N<1$ since the optimal squeezed strategy's energy is
less than the classical one, and the quadratic gain would favor the
classical strategies in that case.  \togli{CHECK THIS: initially we
  had got it wrong!!!}  Intuitively, this can be seen by considering
the minimum uncertainty states for which $\Delta A\Delta H=|\<C\>|/2$.
Then
\begin{align}
\Delta\varphi=\Delta A/|\frac\partial{\partial\varphi}\<A\>|=\Delta
A/\|\<C\>|=1/(2\Delta H)
\labell{ll1}\;.
\end{align}
Then, for $\lambda>1$ we get better accuracy since $\Delta H$ is
increased, whereas the opposite happens for $\lambda<1$. (iii)~The
theory above refers to the estimation of $\varphi$ from the
measurement of an observable $A$, but it applies also to POVM
measurements. Indeed a POVM can be extended to a projective POVM
through the Naimark dilation theorem \cite{peresbook} without changing
the measurement outcome statistics.  One can then assign arbitrary
``eigenvalues'' to each element of this POVM to obtain an observable
$A$ to which the above theory applies.  (iv)~As for entanglement-based
quantum metrology \cite{rafalguta,davidov,rafalreview}, the presence
of noise complicates the situation enormously and will be analyzed
elsewhere.

\section{Examples}
We now show some simple examples that refer to different (finite and
infinite-dimensional) Hilbert spaces {\em and} to different
observables.
\subsection{Position measurements}
Consider the situation where we want
to estimate the position $X$. The generator of translation of position
is the momentum $P$, so we choose $H=P$. This Hamiltonian is not lower
bounded so the average energy is infinite for any state: we need to
introduce an energy cutoff, e.g.~by supposing that states have
negligible negative momenta components.  [Alternatively, one could
also consider $H=|P|$ as the Hamiltonian.]  We consider a
position-momentum squeezed state \cite{yuen}, whose wavefunction is a
Gaussian \cite{yuen1}. Introducing (arbitrary) constants $m$ (mass)
and $\omega$ (angular frequency), we can write $X$ and $P$ in terms of
creation and annihilation operators through
$a=X\sqrt{m\omega/2\hbar}+iP/\sqrt{2\hbar m\omega}$. The squeezed
states are eigenstates of $\mu a+\nu a^\dag$ with
$\mu=(\lambda+1)/\sqrt{4\lambda}$ and
$\nu=(\lambda-1)/\sqrt{4\lambda}$. As before, the coherent state is
obtained for $\lambda=1$, yielding the ``standard'' coherent states.
The rmse are \cite{yuen} $\Delta X=\sqrt{\hbar/(2\lambda m\omega)}$
and $\Delta P=\sqrt{m\omega\hbar\lambda/2}$.  Then, assuming
zero-energy ground state, and using the least energetic squeezed and
coherent states $\<P\>\simeq\Delta P$ (the ones with smallest energy
that still have negligible negative energy components), we have
\begin{align}
N=\frac{\<P\>_{sq}}{\<P\>_{cl}}=\frac{\Delta P_{sq}}{\Delta
  P_{cl}}\simeq\sqrt{\lambda}\;\Rightarrow\frac{\Delta X_{sq}}{\Delta
  X_{cl}}=\frac 1{\sqrt{\lambda}}\simeq\frac1{N}
\labell{casoxp}\;,
\end{align}
a quadratic improvement over the classical strategy of repeating $N$
times the optimal coherent state measurement, which gives $\Delta
X_{cl}(N)/\Delta X_{cl}=1/\sqrt{N}$. 

\subsection{Optical interferometry} As a second example, consider
interferometric phase $\phi$ measurements \cite{peggbarnett}. In this
case $H=a^\dag a\equiv\hat N$ and $A$ is whatever observable (or POVM)
is measured.  Since the phase is periodic, the rmse is not an
appropriate uncertainty measure \cite{holevo} unless the uncertainty
is small compared to $2\pi$.  One must resort to Susskind-Glogower
uncertainty relations (SGUR) \cite{phasermp}
\begin{eqnarray}
\Delta\hat N\Delta \hat C\geqslant\frac12\<\hat
S\>\,,\ \Delta\hat N\Delta \hat S\geqslant\frac12\<\hat
C\>\,
\labell{SG}\;,
\end{eqnarray}
where $\hat C=(E_++E_-)/2$ and $\hat S=i(E_+-E_-)/2$ are the
``cosine'' and ``sine'' operators with $E_\pm=\sum_n|n\>\<n\mp1|$ (in
the Fock basis). Both $\hat S$ and $\hat C$ can be used as the
observable $A$ of the theory given above. Indeed, the SGUR can be used
because $\hat S=\frac\partial{\partial\phi}\hat C=i[H,\hat C]$ and
$\hat C=\frac\partial{\partial\phi}\hat S=i[H,\hat S]$.
\cite{phasermp}.  Hence Eq.~(\ref{sq1}) can be replaced by
$\Delta\phi=\Delta C/|\<S\>|=\Delta S/|\<C\>|$ and we can write SGUR
as $\Delta\hat N\Delta\phi\geqslant\frac12$ (meaningful if
$\Delta\phi\ll2\pi$, with an appropriate choice of boundaries far from
the average $\phi$\togli{check this}). This implies that the minimum
uncertainty squeezed and coherent states for SGUR can be employed in
the theory presented here to get a quadratic improvement for phase
sensing. (The proof is basically identical to the one for $X$ and $P$
presented below, since $\phi$ and $\hat N$ can be considered as
conjugate variables in the above regime, where $\<\phi\>$ is basically
null at the boundaries.)\togli{CHECK THIS} These states were
determined in \cite{jackiw}, but unfortunately they seem to have no
physical relevance, although they can be approximated in particular
regimes, e.g.~\cite{yama,knight}. The usual coherent states
$|\alpha\>$, eigenstates of the annihilation $a$, are not
minimum uncertainty states for SGUR \cite{jackiw}, although they 
approximate them for large average photon number \cite{phasermp}.

The prototypical squeezed-light interferometric measurement is the one
proposed by Caves \cite{caves2}. While the original proposal is not
optimal \cite{rafalreview}, one with a modified detection strategy is:
it has a phase uncertainty $\Delta\phi_{sq}\simeq1/\<a^\dag a\>$ at
the optimal working point \cite{smerzi}.  In contrast, coherent states
$|\alpha\>$ can only achieve the standard quantum limit
$\Delta\phi_{cl}\simeq 1/\sqrt{\<a^\dag a\>}$, so the classical
strategy that optimally employs the energy resources, i.e.~the one for
which $\<a^\dag a\>\simeq\Delta H$\togli{Missing a factor of $\zeta$
  here (and also below)?!  Check this: yes, but we don't care}, is the
one that employs coherent states with $\<a^\dag a\>=1$ since the
coherent state's Poissonian statistic implies that $\<a^\dag
a\>=\Delta H^2$. Then a quadratic gain over {\em this} strategy (when
repeated $N$ times) follows: the energy of the repeated strategy is
$\<a^\dag a\>_N=N$, so that
$\Delta\phi_{sq}/\Delta\phi_{cl}(N)=\sqrt{N}/\<a^\dag a\>=1/\sqrt{N}$
or, when comparing with the single-shot optimal classical strategy,
$\Delta\phi_{sq}/\Delta\phi_{cl}=1/\<a^\dag a \>=1/N$.

As a further example of phase estimation that is optimal only in
certain regimes, consider quadrature squeezing for phase estimation
 (quadrature squeezing is different from SGUR squeezing).
Namely, estimate phase shifts $\phi$ generated by $H=a^\dag a$, by
measuring the quadrature $P=i(a^\dag-a)/\sqrt{2}$. Then
$\Delta\phi=\Delta P/|\<X\>|$, with $X=(a+a^\dag)/\sqrt{2}$ since
$\frac\partial{\partial\phi}P=i[a^\dag a,P]=-X$. For a quadrature
squeezed displaced state we find $\<X\>=\sqrt{2}Re(\alpha)$, and
$\Delta P=e^{-\xi}/\sqrt{2}$, where $\alpha$ and $\xi$ are the
displacement and squeezing parameters and where we choose real $\xi$
(the only interesting case here). Since
$\<H\>=|\alpha|^2+\sinh^2|\xi|$ \cite{dantas}, it is clear that the
good strategies (the ones which do not waste energy) are the ones
where $\alpha_{sq}$ and $\alpha_{cl}$ are real. For these,
\begin{eqnarray}
&\frac{\Delta\phi_{sq}}{\Delta\phi_{cl}}=\frac{\Delta
  P_{sq}}{\Delta
  P_{cl}}\frac{|\<X\>_{cl}|}{|\<X\>_{sq}|}=
\frac{e^{-\xi}|\alpha_{cl}|}{|\alpha_{sq}|}
\labell{sqq}\;,
\\&
N=\frac{\<H\>_{sq}}{\<H\>_{cl}}=
\frac{\sinh^2|\xi|+\alpha_{sq}^2}{\alpha_{cl}^2}
\simeq{\frac12e^{2|\xi|}+\alpha_{sq}^2}
\labell{sqqq},
\end{eqnarray}
where the last equality holds for large squeezing $|\xi|\gg1$ and for
the optimal classical strategy that, again, is for
$\<a^\dag a\>=\Delta H=\Delta H^2={\alpha_{cl}^2}=1$. Since
$x^2+y^2\geqslant 2xy$ for any real $x,y$, we find 
\begin{eqnarray}
\frac{\Delta\phi_{sq}}{\Delta\phi_{cl}}=\frac
1{e^{\xi}|\alpha_{sq}|}\geqslant\frac
2{e^{2|\xi|}/2+\alpha_{sq}^2}=\frac 2{N},\labell{ll}
\end{eqnarray}
if we choose $\xi>0$ (i.e.~$P$-direction squeezing). The inequality
(\ref{ll}) becomes an equality for $\alpha_{sq}^2=\frac12e^{2|\xi|}$,
namely if we devote half of the energy to squeezing and half to
displacing, which is known to be the best way to allocate the energy
\cite{seth}. So, in the limit of large $P$-direction squeezing, we
have optimality also in this case. Many other optical phase estimation
strategies based on squeezing are known,
e.g.~\cite{vittorio,vittorio1,esperim,lund}.

\subsection{Spin squeezing}
As a final example consider spin squeezing, with $A=J_x$ and
$H=-J_y$\togli{CHECK the sign of $\theta$ with this minus sign, does
  it change?! Checked.}. In this case, not all squeezed states are
eigenstates of $L(\lambda)$ since $C$ is not strictly positive or
negative \cite{trifonov}. Thus, instead of limiting ourselves to the
eigenstates of $L(\lambda=1)$, we will employ the $su(2)$ coherent
states \cite{perelomov} which are a larger class \cite{trifonov}.
These are defined as $|\beta\>\equiv\exp(\beta
J_+-\beta^*J_-)|j;-j\>$, with $J_\pm\equiv J_x\pm iJ_y$,
$\beta\in{\mathbb{C}}$, and $|j;-j\>$ the lowest-weight eigenvector of
$J_z$. The latter state is the only eigenstate of
$L(\lambda=1)=J_x-iJ_y\equiv J_-$ (this is the reason for the minus
sign in the definition of $H$, without which we would obtain the
equivalent coherent state class originating from $|j;+j\>$).  The
states $|\beta\>$ have $\Delta J_x^2=j(1-\sin^2\theta\cos^2\phi)/2$,
$\Delta H^2=\Delta J_y^2=j(1-\sin^2\theta\sin^2\phi)/2$,
$\<-J_y\>-E_0=j(1+\sin\theta\sin\phi)$, and
$|\<[J_x,J_y]\>|=|\<J_z\>|=j|\cos\theta|$, where $\theta$ and $\phi$
are defined as $\beta=-e^{i\phi}\tan(\theta/2)$ \cite{perelomov}.  All
these coherent states give rise to a ``too much energy'' strategy,
since $\<H\>_{cl}-E_0>\Delta H$ (except for the case of $|\beta\>$
eigenstate of $J_y$ which is obviously useless for estimation). Thus,
we need to define $N=(\<H\>_{sq}-E_0)/\Delta H_{cl}$ in order to count
only the energy that is actually employed in the estimation in the
classical strategy. The precision achievable in estimating a rotation
by an angle $\varphi$ around the $y$ axis is $\Delta\varphi=\Delta
J_x/|\<J_z\>|=\sqrt{\frac{1-\sin^2\theta\cos^2\phi}{{2j}\cos^2\theta}}
\geqslant \frac1{\sqrt{2j}}$, where the last inequality becomes an
equality on the eigenstates of $J_z$ (i.e.~for $\theta=0,\pi$) as
expected. The squeezed strategy should employ squeezed spin states
\cite{spinsq} with reduced fluctuations in $J_x$, namely $\Delta
J_x=1/\sqrt{2}$, $\Delta J_y=j/\sqrt{2}$ for which $\<-J_y\>+j=j$.
This is a ``good strategy'', since $\<H\>_{sq}-E_0=j\simeq\Delta
H_{sq}=j/\sqrt{2}$.  It achieves the Heisenberg squeezing bound
precision $\Delta\varphi\simeq 1/j$ \cite{spinsq}.  The comparison
between the classical and the squeezed strategies follows:
\begin{align}
&N=(-\<J_y\>_{sq}+j)/{\Delta J_y}_{cl}=j/\sqrt{j/2}=\sqrt{2j}
\nonumber\;\\&
\Rightarrow{\Delta\varphi_{sq}}/{\Delta\varphi_{cl}}=\sqrt{2j}/j=
2/{N},
\labell{ennesp}
\end{align}
again a quadratic improvement (apart from a constant of order one)
over the ${\Delta\varphi_{cl}(N)}/{\Delta\varphi_{cl}}=1/\sqrt{N}$
precision obtained by repeating the classical strategy $N$ times.
Note that in all spin-squeezing literature $N$ is defined differently,
as the number of elementary spin-$1/2$ particles equivalent to a
$j$-spin system \cite{spinsq}. Thus elsewhere spin squeezing
is analyzed in terms of the entanglement among these particles, using
the entanglement-based theory of quantum metrology \cite{qmetr}.

\section{New protocols}
We now detail the step-by-step procedure that one can employ to devise
a new metrology protocol based on squeezing, employing the results of
our paper: \begin{enumerate}
\item Identify the Hamiltonian $H$ that generates the translations of
  the parameter of observable that we want to estimate. Identify the
  observable $A$ that we can employ in the estimation. Note that $A$
  is in general not unique (as shown in the above examples).
\item Solve the eigenvalue equation for the non-Hermitian operator
  $\lambda A+iH$ to find the squeezed states of the system (the
  intelligent states) \cite{trifonov}.
\item Use all the energy available to the estimation procedure to
  prepare a {\em single} probe in such squeezed state.
\item Let the probe evolve with the Hamiltonian $H$ and then measure
  the observable $A$.
\item The results of our paper guarantee that the outcome will have an
  error on the parameter or observable to be estimated that scales as
  $1/N$, where $N$ is the number of classical (coherent) probes that
  could be prepared using the same energy.  These classical probes
  would give an error that scales as the standard quantum limit
  $1/\sqrt{N}$ because of the central limit theorem. Namely, the
  squeezed strategy described here has a quadratic gain.
\end{enumerate}
\section{Conclusions}

In conclusion, we have proposed the general theory of quantum
metrology that is focused on squeezing, instead of on entanglement.
Our framework applies to arbitrary quantum systems, since it uses the
general theory of squeezing. We have also provided various examples of
applications of such theory. We believe that this theory will be of
use to theoreticians to develop new protocols, rather than to
experimentalists.

\vskip 1\baselineskip We acknowledge funding from Unipv, ``Blue sky''
project - grant n.~BSR1718573 and the ATTRACT project funded by
European Union's Horizon 2020 research and innovation programme under
grant agreement No 777222.


\end{document}